\documentstyle[prl,aps]{revtex}
\input epsf
\begin{document}
\draft
%
%%%%%%%%%%%%%%%%%%%%%%%%%%%%%%%%%%%%%%%%%%%%%%%%%%%%%%%%%%%%%%%%%%%%%
%
\title{
Which Chiral Symmetry is Restored in High Temperature QCD?
}
%%%%%%%%%%%%%%%%%%%%%%%%%%%%%%%%%%%%%%%%%%%%%%%%%%%%%%%%%%%%%%%%%%%%%%
%              Compressed author list
%%%%%%%%%%%%%%%%%%%%%%%%%%%%%%%%%%%%%%%%%%%%%%%%%%%%%%%%%%%%%%%%%%%%%%
\author{
Claude Bernard$^1$, Tom Blum$^2$, Carleton DeTar$^3$, 
Steven Gottlieb$^4$, Urs M.~Heller$^5$, James E.~Hetrick$^1$,
K.~Rummukainen$^6$, R.~Sugar$^7$, D.~Toussaint$^8$, Matthew Wingate$^9$
}
\address{
%vspace{5mm}   % for short form
$^1$Washington University, St.~Louis, Missouri 63130, USA\\
$^2$Brookhaven National Laboratory, Upton, New York 11973-5000, USA\\
$^3$University of Utah, Salt Lake City, Utah 84112, USA\\
$^4$Indiana University, Bloomington, Indiana 47405, USA\\
$^5$SCRI, The Florida State University, Tallahassee, 
   Florida 32306-4052, USA\\
$^6$Universit\"at Bielefeld, D-33615 Bielefeld, Germany\\
$^7$University of California, Santa Barbara, California 93106, USA\\
$^8$University of Arizona, Tucson, Arizona 85721, USA\\
$^9$University of Colorado, Boulder, Colorado 80309, USA
}
%%%%%%%%%%%%%%%%%%%%%%%%%%%%%%%%%%%%%%%%%%%%%%%%%%%%%%%%%%%%%%%%%%%%%%
%              Long-form author list
%%%%%%%%%%%%%%%%%%%%%%%%%%%%%%%%%%%%%%%%%%%%%%%%%%%%%%%%%%%%%%%%%%%%%%
%\author{Claude Bernard and James E.~Hetrick }
%%
%\address{
%Department of Physics, Washington University, 
%St.~Louis, Missouri 63130, USA
%}
%%
%\author{Tom Blum}
%%
%\address{
%Brookhaven National Laboratory, Upton, New York 11973-5000, USA
%}
%%
%\author{Matthew Wingate}
%%
%\address{
%Physics Department, University of Colorado, Boulder, Colorado 80309,
%USA }
%%
%\author{Carleton DeTar}
%%
%\address{
%Department of Physics, University of Utah, 
%Salt Lake City, Utah 84112, USA
%}
%%
%\author{Steven Gottlieb}
%}
%%
%\address{
%Physics Department, Indiana University, Bloomington, Indiana 47405, 
%   USA
%}
%%
%\author{Urs M.~Heller}
%%
%\address{
%SCRI, The Florida State University, Tallahassee, Florida 32306-4052, 
%   USA
%}
%%
%\author{Kari Rummukainen}
%%
%\address{Fakult\"at f\"ur Physik, Universit\"at Bielefeld, 
%D-33615 Bielefeld, Germany}
%%
%\author{Robert L.~Sugar}
%%
%\address{
%Department of Physics, University of California, 
%Santa Barbara, California 93106, USA
%%
%\author{Douglas Toussaint}
%%
%\address{
%Department of Physics, University of Arizona, Tucson, Arizona 85721, 
%   USA
%}
%}
%
%%%%%%%%%%%%%%%%%%%%%%%%%%%%%%%%%%%%%%%%%%%%%%%%%%%%%%%%%%%%%%%%%%%%%%
%
\date{\today}
\maketitle
\begin{abstract}
Sigma models for the high temperature phase transition in quantum
chromodynamics (QCD) suggest that at high temperature the
$SU(N_f)\times SU(N_f)$ chiral symmetry becomes exact, but the
anomalous axial $U(1)$ symmetry need not be restored.  In numerical
lattice simulations, traditional methods for detecting symmetry
restoration have sought multiplets in the screening mass spectrum.
However, these methods were imprecise and the results, so far,
incomplete.  With improved statistics and methodology, we are now able
to offer evidence for a restoration of the $SU(2)\times SU(2)$ chiral
symmetry just above the crossover, but not of the axial $U(1)$ chiral
symmetry.
\end{abstract}
%
%%%%%%%%%%%%%%%%%%%%%%%%%%%%%%%%%%%%%%%%%%%%%%%%%%%%%%%%%%%%%%%%%%%%%%%
\section{Introduction}

A high temperature phase transition from a deconfined quark plasma to
a confined phase is thought to have occurred as the early Universe
cooled.  This phenomenon is under investigation in high energy
heavy-ion collisions.  Through numerical simulations of quantum
chromodynamics (QCD) we hope to gain an understanding of the
qualitative and quantitative characteristics of this phase transition.
The phase transition (perhaps only a crossover at physical quark
masses) is associated with the spontaneous breaking of the chiral
symmetry and formation of chiral condensates. Sigma models suggest
that in the limit of zero up and down quark masses, the $SU(2)\times
SU(2)$ chiral symmetry is exact in the high temperature phase
\cite{ref:chiral}, and a phase transition separates it from a cold
phase in which this symmetry is spontaneously broken.  The gauge
anomaly, present at low temperature, may persist at high temperature,
however, breaking the $U(1)$ axial symmetry at all temperatures.

Early efforts to detect symmetry restoration looked for chiral
multiplets in the screening mass spectrum \cite{ref:screening}. For
example, the following channels are related according to the indicated
symmetries:
\begin{displaymath}
\begin{array}{cll}
            & \leftarrow SU(2)\times SU(2) \rightarrow \\
 \uparrow   \\
 U(1)_A     & \ \ \ f_0   & \pi \\
 \downarrow & \ \ \ \eta  & a_0 
\end{array}
\end{displaymath}
The screening mass spectrum is found from the space-like hadron
propagators.  The restoration of the $SU(2)\times SU(2)$ symmetry
requires a degeneracy between the lowest pion screening mass and that
of its chiral partner, the $J^P = 0^+$, $I=0$ $f_0$ meson (also known
as the $\sigma$).  The determination of the $f_0$ screening mass
through numerical simulation is complicated by the presence of
quark-line disconnected graphs.  Computing them requires an expensive
determination of the quark propagator from multiple origins.  In early
simulations, therefore, it was common to keep only connected graphs.
This practice, applied to the $f_0$, results instead in a
determination of the screening mass for the $J^P = 0^+$ $I = 1$ $a_0$
meson (also known as the $\delta$)
\cite{ref:shuryak}.  This meson is the axial $U(1)$ chiral partner of
the pion.  Thus a degeneracy in the $\pi$ and $a_0$ screening masses
would imply a suppression of the gauge anomaly and a partial
restoration of the axial $U(1)$ symmetry, but does not test
restoration of the $SU(2) \times SU(2)$ symmetry.

New simulations with large data samples make it possible to revisit
the question of which symmetry is restored
\cite{ref:eos,ref:nf2_thermo}. Further statistical improvement can be
obtained by studying the susceptibilities related to the propagators,
rather than just the screening masses: for example, from the pion
susceptibility
\begin{equation}
  \chi_\pi = \int d^4r \left\langle \pi(0)\pi(r)\right\rangle
\end{equation}
and the related susceptibilities, $\chi_{f_0}$ and $\chi_{a_0}$, we
can define two order parameters
\begin{equation}
  \chi_{SU(2)\times SU(2)} = \chi_\pi - \chi_{f_0} 
  \ \ \ \mbox{and} \ \ \
  \chi_{U(1)} = \chi_\pi - \chi_{a_0}.
  \label{eq:order_param}
\end{equation}
Restoration of either symmetry requires that the corresponding order
parameter vanish.

We use the staggered fermion scheme.  This scheme breaks all but one
generator of chiral $SU(4) \times SU(4)$.  The full symmetry is
expected to be recovered in the continuum limit.  The one surviving
generator, however, can be used to explore symmetry restoration at the
phase transition at nonzero lattice spacing.  The staggered fermion
treatment of the axial $U(1)$ symmetry is less satisfactory.  That
symmetry, formulated in the conventional manner, is broken explicitly
on the lattice.  It, too, is expected to be recovered in the continuum
limit.  Since our analysis treats only one lattice spacing, namely $a
\approx 1/(6T_c)$, further study will be required to distinguish
between effects of the lattice approximation and continuum effects of
the gauge anomalies.

A preliminary report of our results was presented at Lattice '96
\cite{ref:nf2_thermo}. A number of other groups have also taken up
this question and have also reported preliminary
results\cite{ref:chandra-christ,ref:boyd,ref:lagae}.

\section{Formalism and Computation}

We simulate the $N_f$-flavor staggered fermion action with the
standard partition function at temperature $T$ on a hypercubic
Euclidean lattice with spacing $a$, quark matrix $M(U,m_q)$, quark mass
$m_q$, and gauge link matrices $U$\cite{ref:Montvay}:
\begin{equation}
   Z = e^{-VF(T,am_q)/T} = \int[dU]\exp[-S_g(U)] [\det M(U,m_q)]^{N_f/4}.
 \label{eq:ensemble}
\end{equation}
As is well known, the fermion determinant can be expressed as $ \det
M(U,m_q) = \det[D^2 + (2am_q)^2] $, where the latter determinant is
taken on the even lattice sites only and $D^2$ is the square of the
fermion hopping matrix.  Thus the free energy is manifestly even in
the quark mass.

We will be concerned with a variety of susceptibilities related to the
singlet chiral order parameter,
\begin{equation}
 \left\langle f_0\right\rangle \equiv 
    \left\langle \bar\psi\psi\right\rangle = 
    \partial F(T,m_q)/\partial m_q = 
    TN_fa/2V \left\langle\mathop{\rm Tr} M^{-1}\right\rangle,
\end{equation}
where the expectation values are defined on the ensemble
(\ref{eq:ensemble}).  The associated susceptibility is
\begin{equation}
  \chi_{f_0} = \partial \left\langle f_0\right\rangle/\partial m_q
             = \int d^4x[ \left\langle f_0(0) f_0(x)\right\rangle -
                \left\langle f_0(0)\right\rangle^2]
             = \chi_{\rm conn} + \chi_{\rm disc}
\end{equation}
The quark-line connected and disconnected contributions are
\begin{equation}
   \chi_{\rm conn} 
      = TN_fa^2/V \left\langle\mathop{\rm Tr} M^{-2}\right\rangle
  \ \ \ \mbox{and} \ \ \
   \chi_{\rm disc} = T/V\left[\left\langle (aN_f/2 
           \mathop{\rm Tr} M^{-1})^2\right\rangle
      - \left\langle aN_f/2\mathop{\rm Tr} M^{-1}\right\rangle^2\right]
\end{equation}
It can be seen from this result that the disconnected contribution to
the susceptibility is just proportional to the ``configuration
variance'' of $\left\langle f_0\right\rangle$, that is
$
  \chi_{\rm disc} = V/T \left[\left\langle f_0^2\right\rangle 
      - \left\langle f_0\right\rangle^2 \right].
$

All of our simulations are carried out with two dynamical (sea) quark
flavors.  However, in measuring susceptibilities, we can adjust the
valence flavor number to suit the observable.  If we stick with only
the four flavors forced upon us by fermion doubling in the staggered
fermion scheme, all isospin components of the $a_0$ meson are
generated by a nonlocal fermion bilinear \cite{ref:golterman}.
However, at the expense of increasing the flavor degeneracy to eight,
we can create an $a_0$ analog from a diagonal fermion bilinear
operator.  In any case all such $a_0$ components are expected to be
degenerate in the continuum limit and any of them can be used to test
symmetry restoration.  The susceptibility of the diagonal $a_0$
operator is exactly the connected part of the $f_0$ susceptibility:
\begin{equation}
  \chi_{a_0} = \chi_{\rm conn}.
\end{equation}
%
%%%%%%%%%%%%%%%%%%%%%%%%%%%%%%%%%%%%%%%%%%%%%%%%%%%%%%%%%%%%%%%%%%%%%%
%   Figures for short-form
%%%%%%%%%%%%%%%%%%%%%%%%%%%%%%%%%%%%%%%%%%%%%%%%%%%%%%%%%%%%%%%%%%%%%%
\vspace*{-8cm}
\begin{center}
\begin{tabular}{lr}
\hskip-10mm
\begin{minipage}[t]{7cm}
\begin{figure}
\vbox{\vskip-20mm
\epsfxsize=3.0in \epsfbox[0 0 4096 8192]{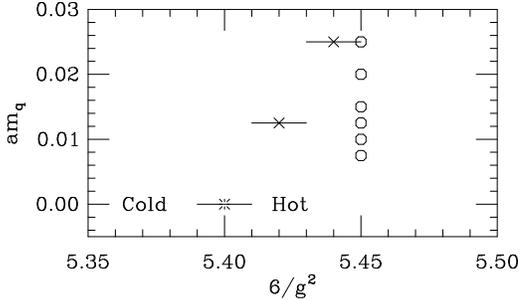}}
\caption{\label{fig:phase_diag} 
Phase diagram for the standard $SU(3)$ Wilson gauge plus two-flavor
staggered fermion action showing the approximate $N_t = 6$ crossover
location (crosses and burst) as a function of gauge coupling $6/g^2$
and quark mass $am_q$.  Data sample points are indicated by octagons.  }
\end{figure}
\end{minipage}
&
\hskip10mm
\begin{minipage}[t]{8cm}
\begin{figure}
\vbox{
\epsfxsize=6.0in \epsfbox[0 0 8192 8192]{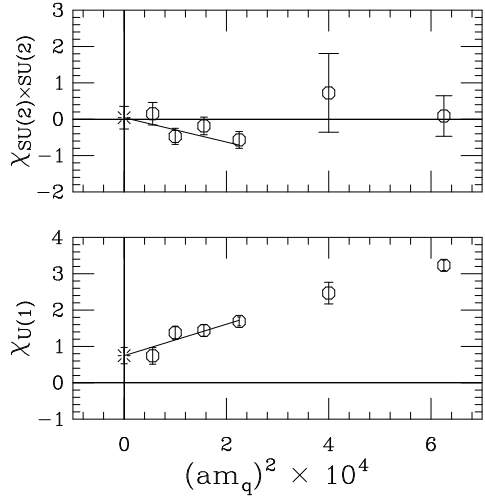}}
\caption{\label{fig:chiral} 
Chiral order parameters extrapolated in quark mass squared.
}
\end{figure}
\end{minipage}
\end{tabular}
\end{center}
%%%%%%%%%%%%%%%%%%%%%%%%%%%%%%%%%%%%%%%%%%%%%%%%%%%%%%%%%%%%%%%%%%%%%%
%
We measure this susceptibility directly from the connected part of the
$f_0$ correlator: $\chi_{\rm conn} = \int d^4x
\left.\left\langle f_0(0)f_0(r)\right\rangle\right|_{\rm conn}$, 
while Chandrasekharan and Christ measure it by taking the derivative
of $\left\langle f_0\right\rangle$ with respect to the valence quark
mass \cite{ref:chandra-christ}. Finally, a well-known Ward identity
relates the pion susceptibility to the chiral order
parameter\cite{ref:toolkit}:
\begin{equation}
   \chi_\pi 
    = N_fTa^2/V\left\langle\mathop{\rm Tr}(M^\dagger M)^{-1}\right\rangle
    = \left\langle f_0\right\rangle/(2m_q).
\end{equation}

In practice we measure the order parameters (\ref{eq:order_param})
through
\begin{equation}
  \chi_{SU(2)\times SU(2)} = \left\langle f_0\right\rangle/(2m_q)
    -  \chi_{\rm conn} - \chi_{\rm disc}
  \ \ \ \mbox{and} \ \ \
  \chi_{U(1)} = \left\langle f_0\right\rangle/(2m_q) - \chi_{\rm conn}
\end{equation}

The simulation consisted of a subset of configurations generated in an
extensive study of the equation of state for $N_t = 6$ and $N_f = 2$
at $6/g^2 = 5.45$ and quark masses $am_q = 0.0075$, $0.01$, $0.0125$,
$0.015$, $0.02$, and $0.025$ \cite{ref:eos,ref:nf2_thermo}. This
parameter range lies in the high temperature phase slightly above the
phase transition, as illustrated in Fig.~\ref{fig:phase_diag}, and was
selected to permit an extrapolation of the measured quantities to zero
quark mass in the high temperature phase.  The simulation sample at
each mass covered a molecular dynamics time span of at least 2000 time
units with the first 400 omitted.  Measurements were taken at
intervals of at most 50 time units.  The chiral order parameter
$\left\langle f_0\right\rangle \equiv \left\langle
\bar\psi\psi\right\rangle$ was measured using the random source method
\cite{ref:rand_pbp} with 33 random sources.  These measurements, with
care taken to avoid biases inherent in the noisy source technique, in
turn, provided an estimate of $\chi_{\rm disc}$ through the
configuration variance.

\section{Results and Conclusions}

Results are shown in Fig.~\ref{fig:chiral} and table
\ref{tab:suscept}.  We have indicated a linear extrapolation in
$(am_q)^2$.  Because they are closer to the crossover
(Fig.~\ref{fig:phase_diag}), where curvature may be expected, we chose
to exclude the two highest mass points from the fit.  The zero mass
intercepts are
\begin{equation}
  \chi_{SU(2)\times SU(2)}  =   0.04(31) 
  \ \ \ \mbox{and} \ \ \
  \chi_{U(1)}               =   0.75(22)
\end{equation}
with $\chi^2/df = 2.6/2$ and $2.5/2$ respectively.  Fits to all points
gave $\chi_{SU(2)\times SU(2)} = -0.33(20)$ with $\chi^2/df = 5.6/4$
and $\chi_{U(1)} = 0.81(11)$ with 2.7/4.

It is surprising that a fit of the same points to an expression {\em
linear} in $am_q$ gives a result consistent with a zero intercept for
{\em both} order parameters: $\chi_{SU(2)\times SU(2)} = 0.15(38)$
with $\chi^2/df = 1.8/2$ and $\chi_{U(1)} = 0.40(56)$ with 2.4/2.  So
which fit is correct?  As we have emphasized, the free energy is
rigorously even in the quark mass.  In consequence the order
parameters are also even.  Thus if the free energy is analytic at zero
quark mass, a quadratic fit is required.  Now some gauge field
configurations give rise to fermion zero modes or near-zero modes.  In
a two-flavor simulation, those modes contribute terms in
$[(am_q)^2]^{N_f/4} = |am_q|$ to the free energy -- terms linear but
nonanalytic.  Such behavior, if not suppressed by a vanishing
probability for encountering zero modes, would imply a phase
transition or infrared singularity at zero quark mass.  However,
measurements of screening masses for $T > T_c$ give no indication of
infrared singularities for small $am_q$.  A phase transition at zero
quark mass for $T > T_c$ is likewise unexpected in sigma models.

In conclusion, our results are consistent with the sigma model
scenario: a restoration of $SU(2)\times SU(2)$ but not of $U(1)_A$
(approximately $3\sigma$).  Whether the apparent breaking of the axial
$U(1)$ symmetry is a lattice artifact or a consequence of the anomaly
remains to be established by future measurements at smaller lattice
spacing and with improved actions.

We would like to thank Edward Shuryak, Norman Christ, Shailesh
Chandrasekharan, Jac Verbaarschot, and Jean-Francois Lagae for helpful
discussions.  This work was supported by the US DOE and NSF.
Computations were done at the San Diego Supercomputer Center, the
Cornell Theory Center, Indiana University, and the University of Utah
Center for High Performance Computing.

%%%%%%%%%%%%%%%%%%%%%%%%%%%%%%%%%%%%%%%%%%%%%%%%%%%%%%%%%%%%%%%%%%%%%
%   Tables
%%%%%%%%%%%%%%%%%%%%%%%%%%%%%%%%%%%%%%%%%%%%%%%%%%%%%%%%%%%%%%%%%%%%%
\begin{table}
\caption{Susceptibilities and order parameters in lattice units.
\label{tab:suscept}
}
\begin{tabular}{ldddll}
$am_q$ & $\left\langle\bar\psi\psi\right\rangle$ 
  & $\chi_{\rm conn}$ & $\chi_{U(1)}$ & $\chi_{\rm disc}$ 
  & $\chi_{SU(2)\times SU(2)}$\\
\hline
0.0075	& 0.0446(12) & 5.21(17) & 0.74(23) & 0.89(21) & \hspace*{7pt}0.15(31)\\
0.01    & 0.0599(16) & 4.61(9)	& 1.38(18) & 0.91(12) & $-$0.47(22) \\
0.0125  & 0.0724(16) & 4.35(9)	& 1.44(16) & 1.25(18) & $-$0.19(22) \\
0.015   & 0.0885(15) & 4.21(7)	& 1.69(12) & 1.12(20) & $-$0.57(23) \\
0.02    & 0.121(5)   & 3.59(14)	& 2.5(3)   & 3.1(1.0) & \hspace*{7pt}0.7(1.1)\\
0.025   & 0.157(3)   & 3.04(8)	& 3.23(14) & 3.3(5)   & \hspace*{7pt}0.1(6)
\end{tabular}
\end{table}
%
%%%%%%%%%%%%%%%%%%%%%%%%%%%%%%%%%%%%%%%%%%%%%%%%%%%%%%%%%%%%%%%%%%%%%%
%   Figures for long form preprint
%%%%%%%%%%%%%%%%%%%%%%%%%%%%%%%%%%%%%%%%%%%%%%%%%%%%%%%%%%%%%%%%%%%%%%
%\begin{figure}
%\vbox{\vskip-20mm
%\epsfxsize=3.0in \epsfbox[0 0 2048 4096]{phase_diag.ps}}
%\caption{\label{fig:phase_diag} 
%Phase diagram for the standard $SU(3)$ Wilson gauge plus two-flavor
%staggered fermion action showing the approximate $N_t = 6$ crossover
%location (crosses and burst) as a function of gauge coupling $6/g^2$
%and quark mass $am_q$.  Data sample points are indicated by octagons.  }
%\end{figure}
%%
%\begin{figure}
%\vbox{
%\epsfxsize=6.0in \epsfbox[0 0 4096 4096]{x_vs_m2.ps}}
%\caption{\label{fig:chiral} 
%Chiral order parameters extrapolated in quark mass squared.
%}
%\end{figure}
%%%%%%%%%%%%%%%%%%%%%%%%%%%%%%%%%%%%%%%%%%%%%%%%%%%%%%%%%%%%%%%%%%%%%%
\end{document}